# Universal dissolution dynamics of a confined sessile droplet

Saptarshi Basu†, D. Chaitanya Kumar Rao, Ankur Chattopadhyay, Joita Chakraborty

Department of Mechanical Engineering, Indian Institute of Science, Bangalore

We experimentally investigate the dissolution of microscale sessile alcohol droplets in water under the influence of impermeable vertical confinement. The introduction of confinement suppresses the mass transport from the droplet to bulk medium in comparison with the non-confined counterpart. Along with a buoyant plume, flow visualization reveals that the dissolution of a confined droplet is hindered by a newly identified mechanism — levitated toroidal vortex. The morphological changes in the flow due to the vortex-induced impediment alters the dissolution rate, resulting in enhancement of droplet lifetime. Further, we have proposed a modification in the key non-dimensional parameters (Rayleigh number $Ra'$ (signifying buoyancy) and Sherwood number $Sh'$ (signifying mass flux)) and droplet lifetime $\tau'_c$, based on the hypothesis of linearly stratified droplet surroundings (with revised concentration difference $\Delta C'$), taking into account the geometry of the confinements. We show that experimental results on droplet dissolution under vertical confinement corroborate universal scaling relations $Sh' \sim Ra'^{1/4}$ and $\tau'_c \sim \Delta C'^{-5/4}$. We also draw attention to the fact that the revised scaling law incorporating the geometry of confinements proposed in the present work can be extended to other known configurations such as droplet dissolution inside a range of channel dimensions, as encountered in a gamut of applications such as micro-fluidic technology and biomedical engineering.

The droplet dissolution process involves the dissolution of lighter single/multiple sessile droplets into a bulk medium by means of mass transfer across the liquid-liquid interface. The mass transfer from the low-density droplet is governed by the interplay of convection and diffusion. The contribution of convection becomes dominant when large droplets with low density dissolve into a dense bulk liquid. An understanding of convection-dominated transport process gives more insights into the fundamental mechanisms that are responsible for the characteristic behaviours of multiphase fluid systems. Further control and tailoring of such processes may lead to a variety of modifications, that can be implemented in several industrial applications such as chemical waste treatment, separation of heavy metals, food processing, distribution of drugs, biomedical diagnostics to name a few [1-5].

From a broader perspective, droplet dissolution can be regarded as a phenomenon analogous to droplet evaporation, which has been extensively investigated by many researchers over the past few decades [6-11]. The effect of confinement in the evaporation dynamics of sessile droplets has been widely studied and it has been reported that the temporal evolution of the droplets exhibits a universal behavior irrespective of the length of confinement [12, 13]. The influence of confinement on evaporating sessile droplets has also been recently investigated for an array of droplets, where it has been shown that the droplet lifetime is a universal function of the degree of confinement for different droplet geometrical parameters (contact angle and contact radius) and surface wettability [14, 15]. While the evaporation is considered mostly diffusion-driven, the dissolution of a droplet can be governed by either diffusion or buoyancy-driven convection, as reported by Dietrich et al. [16]. A transition value of Rayleigh number ($Ra_t \sim 12$) demarcates the two distinguishing regimes. Among different parameters, concentration gradient and size of the droplet (alcohol) have major contributions in controlling the dissolution rate. The complexity of the problem is further amplified, when a multi-component droplet (water-ethanol) was allowed to dissolve in an oil-rich environment [17]. This article demonstrated dominant solutal Marangoni flow within the droplet. However, in the bulk medium, buoyancy-driven convection was preeminent. Bao et al. [18] examined the array of droplets and how the neighbouring droplets collectively influence the droplet lifetime. Droplets surrounded by adjacent ones took longer time to dissolve, while the droplets at the edge have been found to dissolve faster. The computational study by Chong et al. [19] demonstrated that the presence of multiple droplets could lead to the large suppression of mass flux. Retardation of dissolution rate has also been reported by other researchers; where they found that increased thickness of the liquid layer surrounding the immersed droplet hindered mass diffusion [20, 21]. All these articles show that dissolution rates of a single droplet can be manipulated by introducing some sort of spatial confinement, particularly on the same plane the sessile droplet rests. However, none of the systems studied so far have revealed the dissolution dynamics of a vertically confined droplet.

†Email address for correspondence: sbasu@iisc.ac.in

In this letter, we report how the overall dissolution dynamics is modified by positioning the confinement on top of a single droplet. Confinement was placed vertically over the droplet in such a manner (refer Fig. 1) that the movement of the plume, generated due to buoyancy-driven flow, gets obstructed by the impediment. The experiments are carried out on droplets for four different confinement configurations (by varying distance between the droplet and confinement) and their relative contributions in regulating the dissolution rates are assessed both qualitatively as well as quantitatively. We obtain the velocity and vorticity fields surrounding the droplets and discuss the role of a sustained levitated toroidal vortex on the dissolution dynamics of a confined droplet. We provide, a universal estimate of Sherwood number and droplet lifetime as a function of the geometric parameters of the confinement.

In the present work, experiments were conducted with a sessile pentanol droplet, which was allowed to dissolve in a bulk aqueous medium (see Fig. 1 in supplementary material). The water was kept within a clean acrylic cubic tank (5 cm x 5cm x 5cm) having a volume of 100 $mm^3$. A glass slide, coated with a thin PDMS layer, was used as a substrate and was placed at the bottom of the tank. A single 1.5µl 1-pentanol droplet (properties listed in Table 1. of supplementary material) was deposited on that substrate with the help of a liquid dispensing system (F200 Flowline, Precore Solutions). Later, the confinement (a cylinder with diameter 5 mm) was positioned over the alcohol droplet within 30s of droplet deposition. The confinement was designed in such a manner, that when it was placed concentric to the droplet, the projected area of the confinement would encompass the entire droplet area after spreading (see Fig. 1). The temporal evolution of the dissolving sessile droplet was captured with a charge-coupled device (CCD) camera (Nikon D5600) attached with an adjustable lens (Navitar 8x zoom lens) in presence of a diffused backlight source with a resolution of 6000 x 4000 pixels having an exposure time of 1/60 s. The shadowgraph images were recorded at a rate of one frame per second and post-processed using ImageJ to obtain the evolution of radius and volume of the droplet during dissolution. For visualization of the concentration field and flow surrounding the droplet, the water in the tank was seeded with neutrally buoyant hollow glass particles (3 to 30 µm diameter). The tracer particles were illuminated using a laser diode beam, which produced a semi-telecentric laser line with 15 mm line length, 520 nm wavelength, and 29 mW output power (Schäfter + Kirchhoff GmbH, type 13LT). The line width was constant and the intensity profile was uniform in line direction. The focal plane of the laser was centred at the sessile droplet. The light reflected by the glass particles was visualized by the CCD camera at 24 frames per second and a spatial resolution of 1920 x 1080 pixels. The obtained consecutive image pairs were then processed using particle image velocimetry (PIV) software (PIVlab) with an interrogation window of 64 × 64 pixels (corresponding to ~ 40 µm x 40 µm) to produce time-averaged two-dimensional velocity and vorticity fields. The maximum uncertainty in the velocity and vorticity measurement was ± 2%.

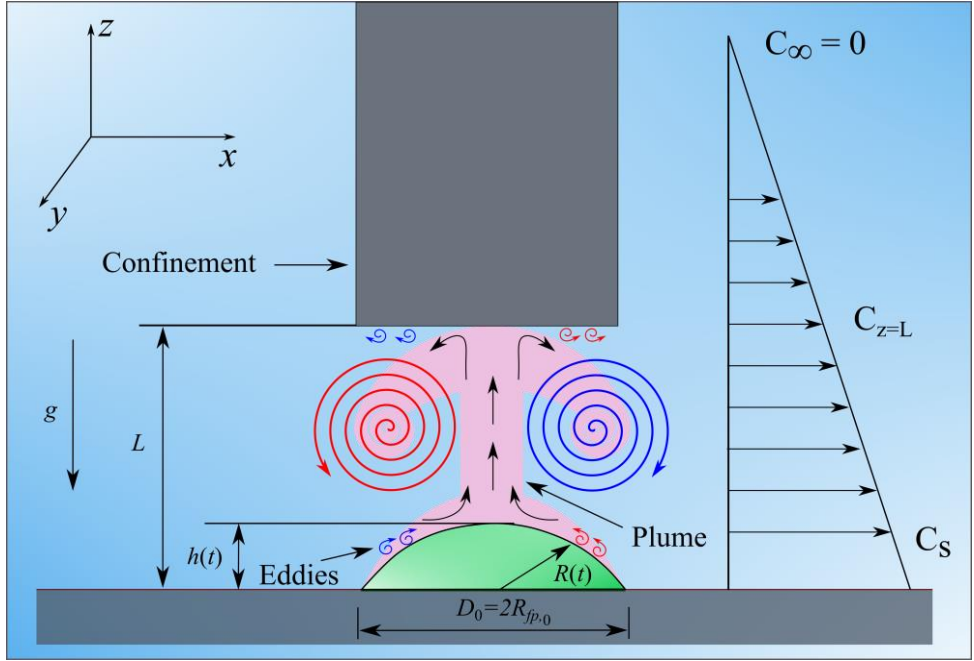

Figure 1. Schematic representation of the dissolution process of the pentanol droplet in water under vertical confinement. The confinement height (*L*) is represented as the distance between the substrate and the edge of confinement, and *D* is the footprint diameter of the droplet. The red and blue spiral structures illustrate the vortex while the small structures shown on the droplet and at the surface of confinement are eddies.

The shadowgraphy images of dissolving single pentanol droplet of the same volume (1.5 µl) in presence of varying confinement distances ($L/D_0$ = 0.75, 1.5, 2.25, 3) were captured throughout the droplet lifetime and were compared with the non-confinement case (which serves as a reference). Figure 1 shows the important characteristic features of the dissolution of a droplet under vertical confinement. Here, *L* is the distance between the substrate and confinement, and *D* is the initial footprint diameter of the droplet.

Since the droplet exhibits stick-jump characteristics during dissolution [22, 23], the equivalent radius (*R*) is expected to give an accurate measurement of droplet size. Figure 2a represents the temporal evolution of the instantaneous droplet radius (*R*) normalized with initial droplet radius ($R_0$) for alcohol droplets for different $L/D_0$ values. Here the instantaneous time (*t*) is normalized with the convective timescale, $\tau_c'$ (discussed later in the manuscript). Similarly, Fig. 2b represents the time evolution of normalized volume, where the droplet volume (*V*) is related to the equivalent radius (*R*) by the following relation

$$V = \frac{4}{3}\left(\pi R^3\right) \tag{1}$$

It is evident from the figure that compared to the non-confined case; the droplets under confinement dissolve at a slower rate. However, it is rather difficult to distinguish the dissolution rate among the confinement configurations with smaller $L/D_0$, which calls for further investigations.

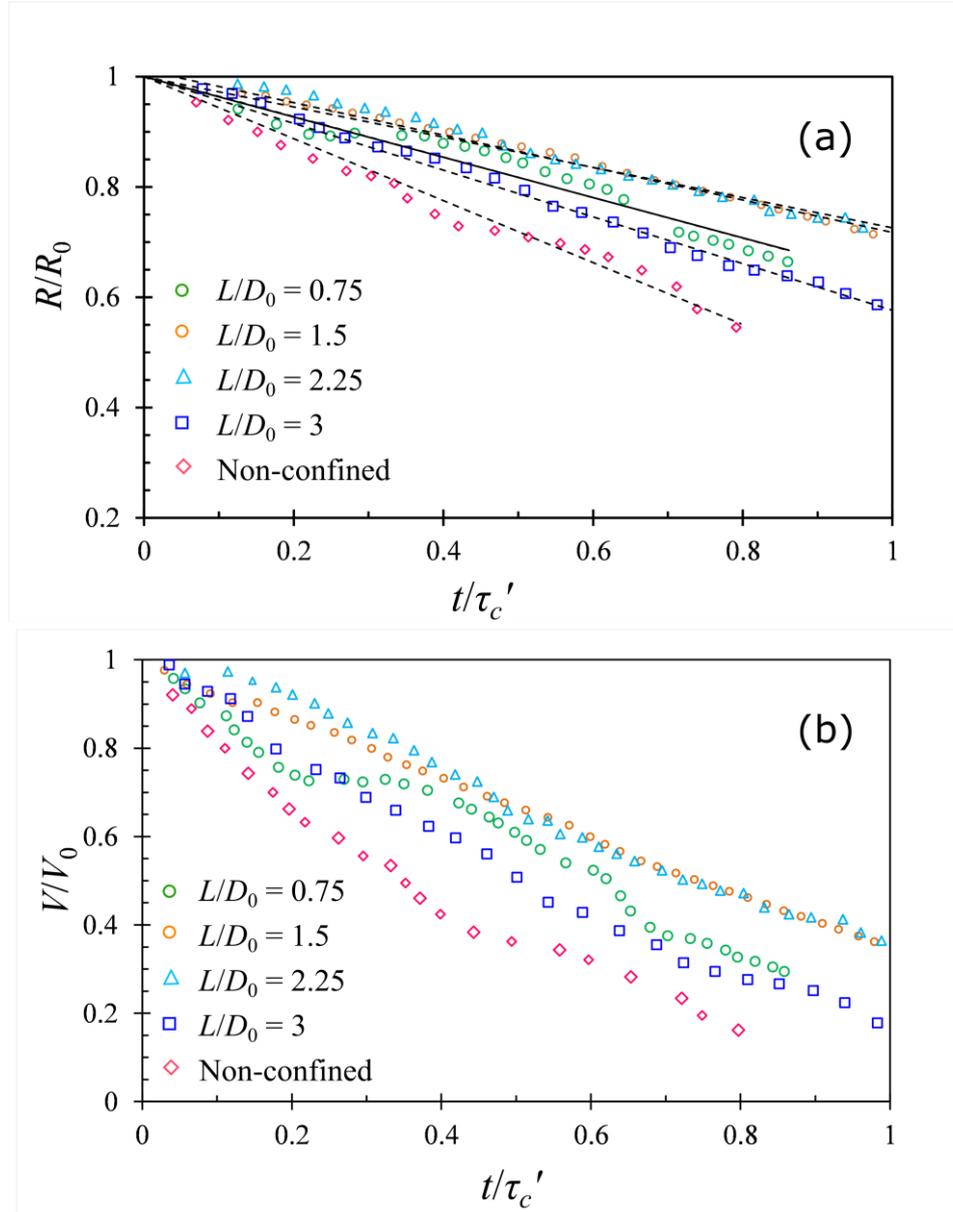

Figure 2. Comparison of temporal evolution of non-confined droplet with four different confinement distances ($L/D_0$ = 0.75, 1.5, 2.25, and 3), depicting (a) Normalized radius as a function of time and (b) normalized volume as a function of time. The instantaneous time is normalized with the convective lifetime of the droplet $\left(\tau_c'\right)$. Here, $R_0$ and $V_0$ are initial droplet radius and droplet volume, respectively. The curves represent the average of at least three different cases with a maximum uncertainty of 3%. The dashed lines in fig. 2(a) represent a linear trend.

The dissolution of the solute (alcohol) droplet into the surroundings (water) results in a relatively low-density alcohol-water mixture compared to the dense bulk medium. This unstable stratified zone aids in the formation of a plume (Fig. 1). Further dissolution of the droplet leads to a long tail of the plume attached to the apex of the droplet until the entire droplet is dissolved.

During the dissolution of droplets under confinement, both the large-scale (low-velocity) and small-scale (high-velocity) structures/flow patterns are noticed. Large scale structures include a toroidal vortex and central plume originating from the droplet apex, while smallest structures are the eddies near the edge of confinement as well as at the interface of droplet surface (Fig. 1). It is important to note that central plume oscillates with time in a stochastic manner, which in turn possibly leads to the deviations in the dissolution rate, particularly for smaller $L/D_0$ configurations. The formation of small structures (counter-rotating eddies) near the confinement and at the droplet interface is due to the boundary layer. The cascade of these counter-rotating eddies is spatially aperiodically produced at the droplet interface, and are carried upward by the induction of the vortex. Several eddies can be seen distributed along the surface of the droplet (Fig. 3b).

Figure 3a and 3b visualizes the velocity and vorticity fields respectively, corresponding to pentanol droplet dissolution in the presence of four different confinement configurations ($L/D_0$ = 0.75, 1.5, 2.25, and 3). The interaction of tip of the plume with the confinement, causes it to turn and eventually develop into a vortex entrapping the surrounding bulk medium. Essentially, the plume is divided into streams after it encounters the confinement, which acts as a bluff body to the flow. A portion of its stream dissolves into the surrounding medium while the rest feeds back to the droplet via the vortex (inset of Fig. 3a, second panel). As a consequence of this feedback, the local concentration gradient in the droplet surrounding decreases with time, leading to an attenuated mass loss rate from the droplet. For all the confinement cases, the vortex remains pronounced and stable through most of the droplet lifetime (see Fig. 3 in the supplementary material for the time-resolved velocity and vorticity fields around pentanol droplets for $L/D_0$ -1.5). In contrast, for a non-confined droplet, the vortex convects away leaving behind only the central plume (See Fig. 2 in supplementary material for the flow field surrounding the non-confined droplet).

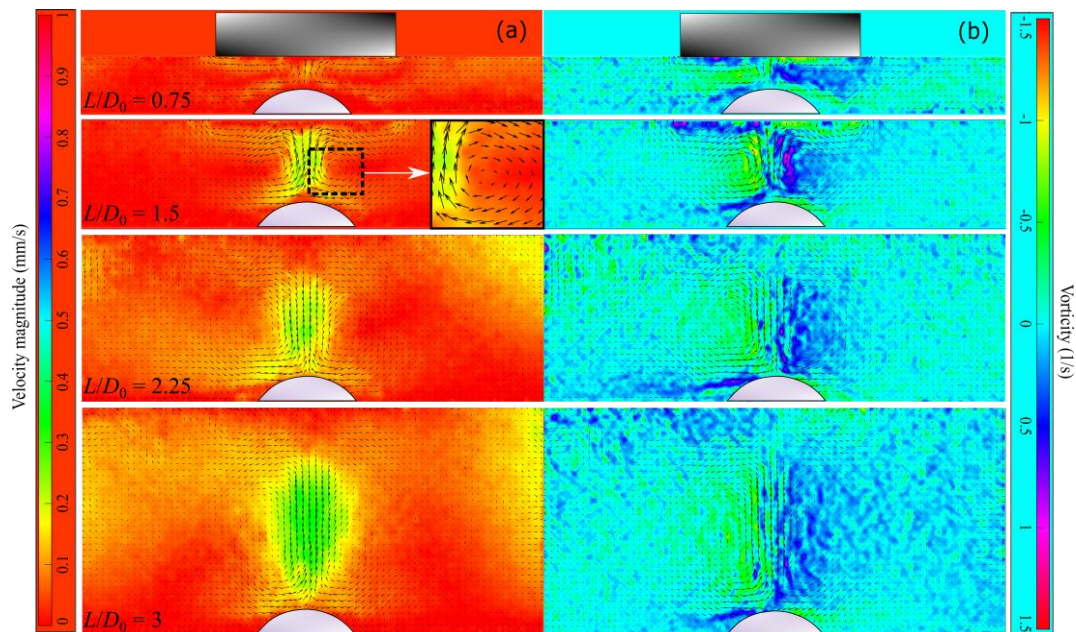

Figure 3. Flow visualization of dissolving droplets under different confinements ($L/D_0$ = 0.75, 1.5, 2.25, and 3) representing (a) velocity magnitude and (b) iso-surface of vorticity taken at 120 s after the deposition of the droplets. The vectors were obtained by taking a mean over 2 seconds (48 frames). The inset in Fig. 3a displays the magnified portion of the vortex. The focal plane of the laser was centered at the droplet.

The diameter of the vortex grows in size (with diameter ~ 0.65 mm, 1.3 mm, 2.3 mm, and 4.2 mm for $L/D_0$ values of 0.75, 1.5, 2.25, and 3 respectively) as the distance between the confinement and droplet is increased. As expected, the time taken for the plume material to complete one rotation and return to its original position is also delayed with $L/D_0$ (with time ~ 55 s, 85 s, 100 s, and 140 s for $L/D_0$ values of 0.75, 1.5, 2.25, and 3 respectively). This indicates that it takes a shorter amount of time for the droplet surroundings to be saturated with the alcohol for smaller $L/D_0$ configurations.

It was observed that the velocity of the plume originating from the apex of the droplet decreases as the confinement is brought closer to the droplet (Fig. 3a). For closer confinement, the plume is shorter with a smaller magnitude of plume velocity ~ 0.12, 0.2, 0.25, and 0.32 mm/s corresponding to $L/D_0$ values of 0.75, 1.5, 2.25. and 3 respectively. For the confinement configuration corresponding to the smallest $L/D_0$ (0.75), one can observe that the central plume is less noticeable since the velocity magnitude of the flow is significantly lower. If the confinement is very close to the droplet, the plume has less distance available for acceleration. Thus, less momentum is available for transfer and, hence, a weak vortex exists in close confinement (Fig 3b, $L/D_0 = 0.75$). As the confinement height ($L$) is raised, larger momentum transfer occurs from the accelerating plume to the vortex, making the vortex stronger. As a result, the velocity of the plume increases (Fig 3a, $L/D_0 = 1.5$), and a stronger iso-surface of vorticity is noticed beside the plume in Fig 3b (second panel, $L/D_0 =1.5$). At this point, the vortex drags the fluid near the droplet with it, similar to an oversimplified Couette-like flow. Accordingly, counter-rotating eddies occur due to viscosity in the boundary layer of the droplet. A further rise in the confinement height causes the strength of the vortex to reach saturation. This is clear from the iso-surface of vorticity in Fig 3b (third panel, $L/D_0 = 2.25$ onwards). This is because, although the plume keeps accelerating, the transferred momentum to the toroidal vortex is further expended in transporting the eddies from the boundary layer of the droplet upward towards the solid surface of the confinement. As the plume moves upwards, due to momentum diffusion, it drags the surrounding quiescent liquid and leads to gradual growth in velocity boundary layer thickness. Moreover, it was also noticed that the velocity magnitude reduces along with the height of the plume (see Fig. 4 in the supplementary material).

The concentration difference along the plane of the plume can be approximated using linear stratification theory. The plume is lighter and after hitting the obstacle discharges horizontally to a comparatively heavier fluid (Fig. 3). Therefore, an obvious gradient of concentration is plausible and with time, the discharge into the bulk leads to the density stratification. It is fair to assume that the medium is linearly stratified (the mathematical part is discussed later). The dissolution timescale is very long while the discharge rate (~mm/sec) is rather quick leading to a quasi-steady linearly stratified medium for majority of the dissolution process.

Now along with visualization, we can quantify that dissolution rate decreases with decreasing $L/D_0$ because convective mass transfer decreases. In this fluid system, mass transfer occurs because of concentration difference, hence a relation between total mass flux and (average) concentration difference is required. We express the results using non-dimensional numbers such that it can be implemented to system independent applications. The non-dimensional number taking care of this is Sherwood number ($Sh$), defined as the ratio of total mass flux to the mass flux by pure diffusion [24]

$$Sh = \frac{\dot{m}_A R}{\delta \Delta C} \qquad (2)$$

where $\dot{m}_A$ is the measured mass flux rate averaged over the entire surface area ($A$) (in kg/m²s) of the droplet and $\delta$ is diffusivity coefficient (m²/s) (properties of pentanol are listen in Table 1 of the supplementary material). As discussed before, it is important to note that buoyant plume fluctuates with time and does not always interact with the confinement throughout the droplet lifetime. Therefore, in the present work, the rate of mass loss ($\dot{m}_A$) from the droplet is evaluated by extracting temporal variation of volume ( Fig. 2b) averaged over the time considering the plume rises from the droplet apex and is in contact with the confinement. The concentration difference ($\Delta C = C_s - C_\infty$) is the difference in the alcohol concentration between the droplet surface location ($C_s$) and far away from the droplet ($C_\infty$). In Eqn. 2, the expression $\frac{\delta \Delta C}{R}$ accounts for the mass flux due to steady diffusion from a spherical alcohol droplet of radius $R$ ($\theta = 90º$, $\theta$ is equilibrium contact angle). Although in the present study, $\theta$ is less than 90º (25º < $\theta$ < 35º); the above relation can be used to calculate $Sh$ as no significant differences are observed for diffusion-limited case in the present range of contact angles [16].

For a non-confined droplet, $C_\infty = 0$, but at the same time, at any finite distance, local concentration will be finite. For any confined case, the concentration at any finite location; say, adjacent to confinement plane (z = L) will be non-zero due to linear stratification (Fig. 1) and needs to be estimated precisely. In order to predict the revised concentration difference ($\Delta C'$), we invoke the linear stratification assumption. The estimation of the local concentrations corresponding to different confinement locations helps us to modify the *Sh* (given in equation 2) accordingly and the modified Sherwood number (*Sh'*) can be expressed as

$$Sh' = \frac{\dot{m}_A R}{\delta \Delta C'} \tag{3}$$

The amended concentration difference is represented by $\Delta C' = C_s - C_{z=L}$, thereby taking into account of the difference in alcohol concentration between the droplet surface location (z = 0) and at the edge of confinement (z = L). The alcohol concentration along the plane of the confinement is $C_{z=L} = C_s \times (h_0 / L)$, where $h_0$ is the droplet height during the start of the dissolution and *L* is the distance between the substrate and the confinement (see Fig. 1).

Similar to Sherwood number (Sh'), revised Rayleigh number (ratio of the buoyant force to the damping force) can be expressed as [24]

$$Ra' = \frac{g \beta \Delta C' R_0^3}{\nu \delta} \tag{4}$$

where *g* is the acceleration due to gravity (m/s$^2$), *β* indicates solutal expansion coefficient (m$^3$/kg) and *v* represents the kinematic viscosity (m$^2$/s). To incorporate the effect of geometric dependence (confinement), we have modified the concentration difference by invoking the linear stratification assumption (discussed earlier).

$$Ra' = Ra \times f(D/L) \tag{5}$$

where the scaling function *f (D/L)* can be represented as

$$f(D/L) = 1 - \frac{h_0}{L} \quad \text{and} \quad h_0 = \alpha D_0 \tan(\theta/2) \tag{6}$$

In eqn. 6, $D_0$ indicates the footprint diameter of the droplet at the inception of dissolution and θ is equilibrium contact angle of pentanol on PDMS substrate (*θ* ~ 30º). Here, *α* is a constant and it is equal to 0.95. The value of α falls well within the limit (0.92-1), proposed by Meric et al [25], who have analyzed the evaporation of water drops by considering the pseudo spherical cap geometry. Figure 4 shows the *Sh'* as a function of *Ra'* for all cases (non-confinement and confined ones), indicating that the data follows *Sh'* ~ *Ra'* $^{1/4}$ scaling. This observation is comparable with the results reported by Dietrich et al. [16] for non-confined droplets. As we increase the *L/D$_0$*, *Sh'* increases with *Ra'* signifying a rise in convection dominated mass transfer (thus faster rate of dissolution). The results imply that, by introducing confinement close to the droplet, the positive concentration gradient (dC/dz) increases, thereby leading to a steady decline in the total mass transfer rate. The solid line represented in Fig. 4 denotes the universal relation, *Sh* ~ *Ra*$^{1/4}$ for convection domination dissolution [24].

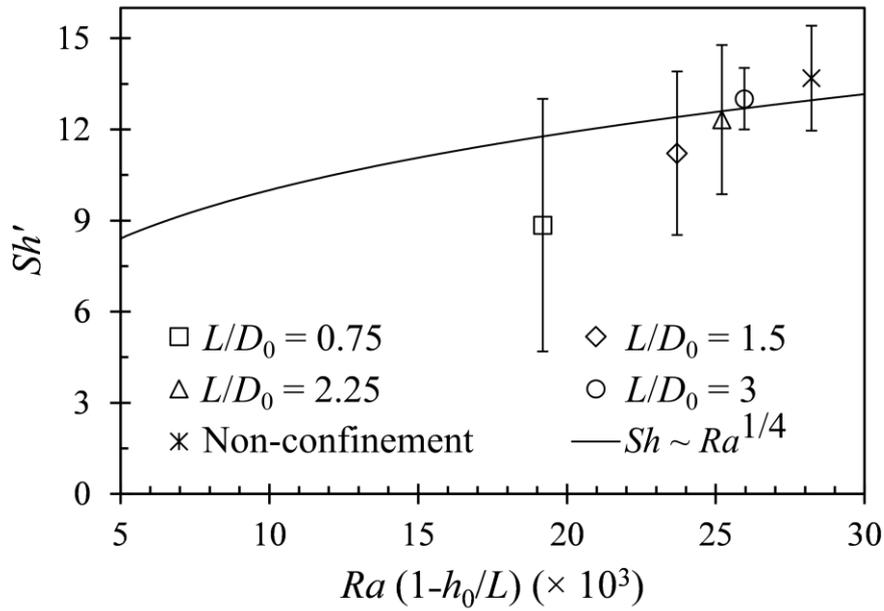

Figure 4. Sherwood number ($Sh'$) as a function of Rayleigh number (Ra) for non-confined and four confinement ratios. The plot shows the average value and the solid line represents $Sh=Ra^{1/4}$. The modified Rayleigh number, $Ra'$ is expressed as a function of geometrical parameters taking linear stratification into consideration, where $h_0$ is the initial droplet height and $L$ is the confinement distance.

As it is established from the ongoing discussion that, the dominant mechanism of mass transfer in the present case is via convection, hence it is important to calculate to the time ($\tau_c'$) taken by the droplet to completely dissolve into the bulk. Although during dissolution, the radius decreases with time, this can be neglected by assuming the process to be quasi-static. The rate of mass loss, $\dot{m}_A \sim -R^2 \rho (dR/dt)$ from the pentanol droplet, is expected to be of the same order of the mass, $\dot{m}_p = A\delta\Delta C' Sh'/R$ carried away by the plume over the entire area (A) of the droplet-bulk interface [16]. We will obtain the following relation by equating both of them

$$\frac{dR}{dt} = -a\left(\frac{g\beta\Delta C'^5 \delta^3}{\nu\rho^4 R}\right)^{1/4} \qquad (7)$$

after integration of $R$ from $R = R_0$ to $R = 0$, and time from $t = 0$ to $t = \tau_c'$ gives the dissolution time,

$$\tau_c' = \frac{4}{5a}\left(\frac{\nu\rho^5 R_0^5}{g\beta\Delta C'^5 \delta^3}\right)^{1/4} \qquad (8)$$

where $R_0$ is the equivalent droplet radius at the beginning of the dissolution process and $\rho$ is the density of pentanol. It turns out from Eqn. 8 that for a given volume of droplet undergoing dissolution in a bulk, the droplet lifetime has inverse dependence on the concentration gradient, i.e, $\tau_c' \propto \Delta C'^{-5/4}$. Here, the material-dependent prefactor 'a' assumes a constant value of 11 for all the cases. The measured (experimental) values of the droplet lifetime for all the confined cases and the non-confinement one is compared against the theoretical one (calculated by eqn. 8), showing a good degree of agreement (Fig. 5).

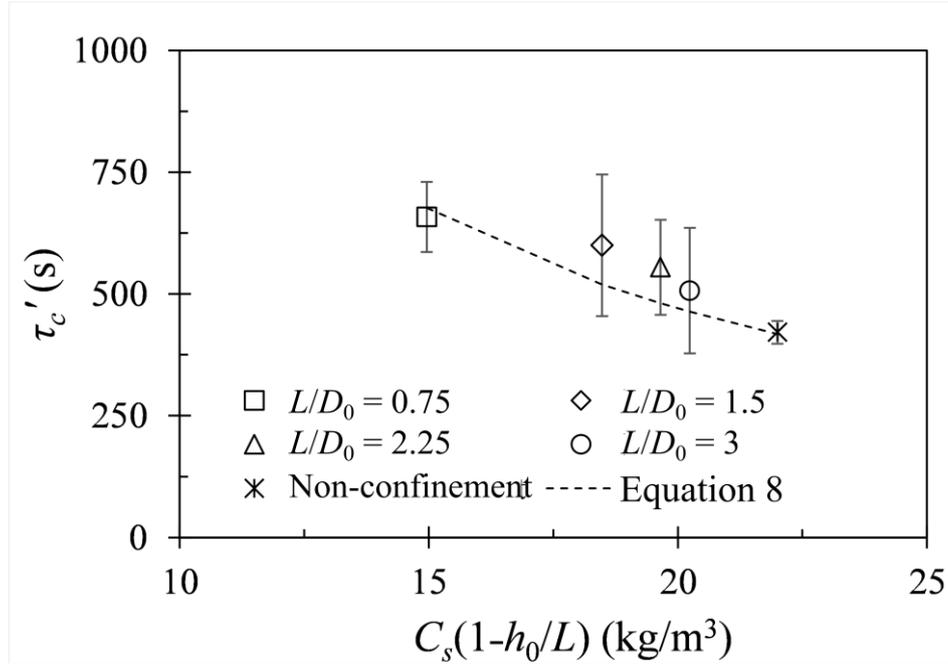

Figure 5. The variation of dissolution time as a function of the concentration difference. The dotted line represents the theoretical dissolution time obtained from equation 7. The modified concentration difference, $\Delta C'$ is expressed as a function of geometrical parameters considering linear stratification.

In summary, in this letter, we have shown that placing vertical confinement over a dissolving droplet is one of the probable means to control its mass flux. The introduction of confinements suppresses the mass flux compared to the non-confined case and delays the time required for the droplet to completely dissolve. In fact, nearer the confinement (smaller $L/D_0$), more is the time taken by the alcohol droplet to dissolve. We report the coupling of the plume and toroidal vortex, where a part of the plume dissolves into the surrounding medium, while the rest feeds back to the droplet via the vortex. Due to this feedback, the local concentration difference surrounding the droplet decreases with time. We formulate a novel method by invoking the linear stratification assumption to estimate local concentration difference and subsequently propose modified non-dimensional numbers. We estimate $Sh'$, $Ra'$, and $\tau_c'$ for different geometrical configurations ($L/D_0$) and they have been found to follow universal relations, $Sh' \sim Ra'^{1/4}$ and $\tau_c' \sim \Delta C'^{-5/4}$. The present work also signifies that, by using the scaling function $\left(1-\dfrac{h_0}{L}\right)$, we can uniquely evaluate the mass transfer properties pertaining to any configurations of vertical confinements, droplet sizes, and properties.

**References**:

1. K. Fukumoto, M. Yoshizawa, and H. Ohno. "Room temperature ionic liquids from 20 natural amino acids." *Journal of the American Chemical Society* 127, no. 8, 2398-2399, (2005).
2. P. Chasanis, M. Brass, and E. Y. Kenig. "Investigation of multicomponent mass transfer in liquid–liquid extraction systems at microscale." *International journal of heat and mass transfer* 53, no. 19-20, 3758-3763, (2010).
3. Z. Lu, A. Rezk, F. Jativa, L. Yeo, X. Zhang, "Dissolution dynamics of a suspension droplet in a binary solution for controlled nanoparticle assembly." Nanoscale.;9(36):13441-8, (2017).
4. A. Jain, and K. K. Verma. "Recent advances in applications of single-drop microextraction: a review." Analytica chimica acta 706, no. 1 (2011): 37-65.
5. D. Lohse and X. Zhang. "Surface nanobubbles and nanodroplets". Review of modern physics, 87, 981-1035 (2015)

6. R. G. Picknett, and R. Bexon, "The evaporation of sessile or pendant drops in still air." J. Colloid Interface Sci. 61 (2), 336-350, (1977).
7. R. D. Deegan, O. Bakajin, T. F. Dupont, G. Huber, S. R. Nagel, & T. A. Witten, "Capillary flow as the cause of ring stains from dried liquid drops." Nature 389 (6653), 827, (1997).
8. N. Shahidzadeh-Bonn, S. Rafai, A. Azouni, and D. Bonn, "Evaporating droplets." J. Fluid Mech. 549, 307-313, (2006).
9. Edwards, A. M. J., Atkinson, P. S., Cheung, C. S., Liang, H., Fairhurst, D. J. &Ouali, F. F. 2018 Density-driven flows in evaporating binary liquid droplets. Phys. Rev. Lett. 121 (18), 184501.
10. Kim, H., Muller, K., Shardt, O., Afkhami, S. & Stone, H. A. 2017 Solutal Marangoni flows of miscible liquids drive transport without surface contamination. Nature Physics 13 (11), 1105.
11. Li, Y., Diddens, C., Lv, P., Wijshoff, H., Versluis, M. & Lohse, D. 2019 Gravitational effect in evaporating binary microdroplets. Phys. Rev. Lett. 122 (11), 114501.
12. Bansal, L., Hatte, S., Basu, S., & Chakraborty, S. (2017). Universal evaporation dynamics of a confined sessile droplet. Applied Physics Letters, 111(10), 101601.
13. Bansal, L., Chakraborty, S., &Basu, S. (2017). Confinement-induced alterations in the evaporation dynamics of sessile droplets. Soft Matter, 13(5), 969-977.
14. Hatte, S., Pandey, K., Pandey, K., Chakraborty, S., &Basu, S. (2019). Universal evaporation dynamics of ordered arrays of sessile droplets. Journal of Fluid Mechanics, 866, 61-81.
15. Pandey, K., Hatte, S., Pandey, K., Chakraborty, S., &Basu, S. (2020). Cooperative evaporation in two-dimensional droplet arrays. Physical Review E, 101(4), 043101.
16. E. Dietrich, S. Wildeman, C. W. Visser, K. Hofhuis, E. Stefan Kooij, H. JW Zandvliet, and D. Lohse. "Role of natural convection in the dissolution of sessile droplets." Journal of fluid mechanics 794 ,45-67, (2016).
17. H. Tan, C. Diddens, A. A. Mohammed, J. Li, M. Versluis, X. Zhang, and D. Lohse. "Microdroplet nucleation by dissolution of a multicomponent drop in a host liquid." Journal of fluid mechanics 870, 217-246, (2019).
18. L. Bao, V. Spandan, Y. Yang, B. Dyett, R. Verzicco, D. Lohse, and X. Zhang. "Flow-induced dissolution of femtoliter surface droplet arrays." Lab on a Chip 18, no. 7 1066-1074, (2018).
19. K. L. Chong, Y. Li, C. Shen Ng, R. Verzicco, and D. Lohse. "Convection-dominated dissolution for single and multiple immersed sessile droplets." Journal of Fluid Mechanics 892 (2020).
20. Q. Xie, and J. Harting. "The effect of the liquid layer thickness on the dissolution of immersed surface droplets." Soft matter 15, no. 32 , 6461-6468, (2019).
21. X. Li, Y. Wang, B. Zeng, Y. Li, H. Tan, H. J. W. Zandvliet, X. Zhang, and D. Lohse, "Entrapment and dissolution of microbubbles inside microwells." Langmuir 34 (36), 10659-10667, (2018).
22. E. Dietrich, E. Stefan Kooij, X. Zhang, H. JW Zandvliet, and D. Lohse. "Stick-jump mode in surface droplet dissolution." Langmuir 31, no. 16 (2015): 4696-4703.
23. X. Zhang, J. Wang, L. Bao, E. Dietrich, R. CA van der Veen, S. Peng, J. Friend, H. JW Zandvliet, L. Yeo, and D. Lohse. "Mixed mode of dissolving immersed nanodroplets at a solid–water interface." Soft Matter 11, no. 10 (2015): 1889-1900.
24. Bejan, A. (2013). *Convection heat transfer*. John wiley & sons.
25. Meric, R. Alsan, and H. Yildirim Erbil. "Evaporation of sessile drops on solid surfaces: Pseudospherical cap geometry." Langmuir 14, no. 7 (1998): 1915-1920.